\documentclass[prX,showpacs,superscriptaddress,twocolumn ]{revtex4}

\usepackage{graphicx}
\usepackage{amsmath}
\usepackage{amsfonts}
\usepackage{amssymb}
\usepackage{epsf}
\usepackage{color}
\usepackage{hyperref}
\newcommand{\bs}{\boldsymbol}

\newcommand{\emma}[1]{#1}

\begin{document}
\title{Exact non-equilibrium solutions of the Boltzmann equation under a
time-dependent external force}

\author{D. Gu\'ery-Odelin}
\affiliation{Laboratoire de Collisions Agr\'egats R\'eactivit\'e,
CNRS UMR 5589, IRSAMC, Universit\'e de Toulouse (UPS), 118 Route de
Narbonne, 31062 Toulouse CEDEX 4, France}

\author{J. G. Muga}
\affiliation{Departamento de Qu\'{\i}mica F\'{\i}sica, Universidad del
Pa\'{\i}s Vasco, Apartado 644, 48080 Bilbao, Spain}
\affiliation{Department of Physics, Shanghai University,
200444 Shanghai, P. R. China}

\author{M.J. Ruiz-Montero}
\affiliation{F´\i sica T\'eorica, Universidad de Sevilla,
Apartado de Correos 1065, E-41080, Sevilla, Spain}

\author{E. Trizac}
\affiliation{Laboratoire de Physique Th\'eorique et Mod\`eles Statistiques (CNRS UMR 8626), B\^atiment 100, Universit\'e Paris-Sud, 91405 Orsay cedex, France}

 \date{\today}

\begin{abstract}
We construct a novel class of exact solutions to the Boltzmann equation,
in both its classical and quantum formulation, for arbitrary
collision laws. When the system is subjected to a specific  external forcing,
the precise form of which is worked out, non equilibrium damping-less solutions
are admissible. They do not contradict the $H$-theorem, but are constructed 
from its requirements. Interestingly, these solutions hold 
for time-dependent confinement. We exploit them, in a reverse-engineering perspective,
to work out a protocol that shortcuts any adiabatic transformation between two equilibrium 
states in an arbitrarily short time-span, for an interacting system.
\emma{Particle simulations of the direct Monte Carlo type fully corroborate the analytical predictions.}
\end{abstract}


\maketitle

More than 140 years after its derivation by J. Maxwell \cite{Maxwell}
and L. Boltzmann \cite{Boltzmann1872}, the so-called Boltzmann equation 
has kept the essence of its original formulation, its predictive power
and its interest. It lies at the heart of the theory of transport in solids,
energy transfer in plasmas, space shuttle aerodynamics, complex flows in
micro-electromechanical systems, neutron transport in nuclear reactors, 
or granular gas dynamics, to name but a few relevant
applications in non equilibrium statistical physics \cite{Cercignani,Kremer}. 
It moreover gradually 
became a thriving branch of mathematics, particularly active in the last 
20 years, see e.g. \cite{Villani02,SaintRaymond} and references therein.	
The Boltzmann equation applies to systems that are rarefied in some
sense, such as an ultracold gas which provides an appropriate 
setting to illustrate the forthcoming 
discussion \cite{pethick1,david99,pethick2,quadupolemonopole0,quadupolemonopole1,quadupolemonopole2}. 

Working on kinetic theory circa 1870 was a leap of faith, impeded by the
controversy pertaining to the atomic nature of matter. It is a great
triumph of Boltzmann to have derived the $H$-theorem, showing that 
the system under scrutiny evolves towards equilibrium, thereby bridging 
microscopic dynamics and macroscopic irreversibility. To this end,
a Lyapunov function was constructed, a non-increasing functional of the 
probability distribution function $f({\bs r},{\bs v},t)$ for finding
gas molecules at position ${\bs r}$ with velocity ${\bs v}$ at a given time $t$. 
It was historically the first
Lyapunov function, and it attributed a direction in time to the Boltzmann
equation. A consequence of the $H$-theorem is that at long times, 
$\log f$ is a collisional invariant, which as such should exhaust 
all independently conserved quantities (momentum, energy in addition to a trivial 
yet relevant constant), so that for like-mass molecules,
$\log f$ should be a linear combination 
of 1, $\bs v$ and $v^2$:
\begin{equation}
f({\bs r},{\bs v},t) \,=\,
\exp\left\{-\alpha -\beta v^2-{\bs\gamma} \cdot {\bs v}
\right\} .
\label{eq:scaling}
\end{equation}
In all generality, $\alpha$, $\beta$ and $\bs \gamma$ are both position 
and time dependent, with the constraint $\beta>0$  \cite{comment3}.
The well known Maxwell-Boltzmann Gaussian form with constant (say vanishing) $\bs \gamma$ 
and a constant inverse temperature $\beta$
is a particular solution. Less known, but nevertheless 
recognized by Boltzmann himself \cite{Boltzmann,Cercignani} is the fact that more
exotic solutions could exist under harmonic 
confinement, that are still of the form (\ref{eq:scaling}) but with a time
dependent kinetic temperature $\propto\beta^{-1}$. These solutions can be envisioned as breathing
modes, where a perpetual conversion of kinetic and potential energy 
operates through a swing-like mechanism and it is essential here that 
the coupling term $\bs\gamma$ be position and time dependent. Another remarkable feature of 
the breathing mode is that it is not restricted to small amplitude oscillations. 
These somewhat non standard solutions were hitherto considered as a side curiosity,
a point of view epitomized by Uhlenbeck, who wrote 
``{\it...for special outside potentials for instance the harmonic potential
$U(r)=\omega r^2/2$ the spatial equilibrium distribution will not be reached
in time. For such special potentials there are a host of special
solutions of the Boltzmann equation (...) where the (coefficients) can
be functions of space and time. (...) They have however only a limited
interest}" \cite{Uhl63}. Uhlenbeck's statement applies to static confinement;
It is our goal here to show the possibility of generalized 
breathing modes for {\em time-dependent forcing}, and to make use 
of these modes to propose a
new kind of gas manipulation on a timescale much shorter than the one dictated 
by the thermodynamical adiabaticity criterion. In doing so, we
put forward a reverse engineering perspective, opposed to the direct approach of Uhlenbeck,
and applying in the quantum realm as well.
Similar protocols have recently been dubbed `shortcuts to adiabaticity' 
\cite{Chen1,reviewsta} in quantum systems \cite{comment}, 
and brought to bear in the realm of transport \cite{TransportDavid}, 
wave-packet splitting or internal state control of single atoms, ions, 
or Bose-Einstein condensates and other many-body systems \cite{Adol}. 
However, in contrast with other phase space manipulation techniques such as the 
Delta Kick Cooling \cite{deltakick}, 
the method proposed here is operational 
for interacting systems and on an arbitrary short time scale.
As a byproduct of the analysis, we uncover for static confinement
new particular potentials allowing for the perpetual non equilibrium 
solutions of the form (\ref{eq:scaling}). Surprisingly, these solutions were missed
by Boltzmann, an omission that propagated ever since in the literature.

The Boltzmann equation hinges on a low density prerequisite which dramatically
simplifies the exact $N$-body dynamics into a non-linear integro-differential
equation for the single particle distribution 
$f({\bs r},{\bs v},t)$. Its rate of change stems from two effects,
free streaming and binary collisions, which translate into the balance
equation \cite{Cercignani,Kremer}
\begin{equation}
\left(\partial_t \, + \, {\bs v} \!\cdot\! {\bs \nabla}_{\bs r} 
+ {\bs F}({\bs r},t) \!\cdot\! {\bs \nabla}_{\bs v} \right) f
\,=\,
I_{\text{coll}}[{\bs v}|f,f] ,
\label{eq:BE}
\end{equation}
where the external (trap) force $\bs F({\bs r},t)$ denotes a position and 
time-dependent field that will be considered conservative:
$\bs F = -{\bs \nabla}_{\bs r} V$. For simplicity, we assume that 
all molecules have the same (unit) mass.
The collisional integral $I_{\text{coll}}$ is a bilinear 
operator acting on $f$, which depends on the precise form of
scattering law considered. We shall not need to specify it further since all
solutions inspected will be of the form (\ref{eq:scaling}) and by virtue
of the $H$-theorem, they identically nullify $I_{\text{coll}}$. 
It is straightforward to check that the equilibrium barometric law 
$f \propto \exp(-2\beta V - \beta v^2)$ is a solution for the Boltzmann 
equation (\ref{eq:BE}). As alluded to above, Boltzmann realized that for 
a harmonic static trap ($V\propto r^2$), more general oscillating solutions of the 
form (\ref{eq:scaling}) were admissible \cite{Boltzmann}. In repeating his argument,
subsequent authors systematically missed other forms of confinement 
that turn out to be compatible with a breathing behavior. 
Our goal is however more general than correcting for that shortcoming,
and and we will explore the venue opened by a suitably chosen
time-dependent trapping, a so far untouched question. To this end,
we introduce
relation (\ref{eq:scaling}) into (\ref{eq:BE}), which leads to
\begin{eqnarray}
&& {\bs \nabla_{\bs r}} \, \beta \,=\, \bs 0 ,
\label{eq:cond1}\\
&& v^2 \partial_t \beta + {\bs v}\cdot\! {\bs \nabla_{\bs r}} (\bs \gamma\cdot\!\bs v) \,=\, 0 ,\quad
\hbox{for all } \bs v ,
\label{eq:cond2}\\
&&\partial_t \bs \gamma + \bs\nabla_{\bs r} \, \alpha + 2 \beta \bs F \,=\, \bs 0 ,
\label{eq:cond3}\\
&&\partial_t \alpha + \bs F \cdot \bs \gamma \,=\, 0.
\label{eq:cond4}
\end{eqnarray}
Any triplet ($\alpha$, $\beta$, $\bs\gamma$) fulfilling 
Eqs.~(\ref{eq:cond1})-(\ref{eq:cond4}) is a solution 
to Eq.~(\ref{eq:BE}), and we of course recover the barometric law
($\alpha = 2 \beta V$, $\beta$ uniform and constant, $\bs\gamma= \bs 0$)
among all possible solutions.

The structure of the system (\ref{eq:cond1})-(\ref{eq:cond4}) constrains the possible form 
of $\bs F = -\bs\nabla_{\bs r} V$, a feature which we now analyze.
We learn from Eqs.~(\ref{eq:cond1})  that
$\beta$ is a sole function of time. In addition, the general solution of 
(\ref{eq:cond2}) can be written
\begin{equation}
\bs \gamma (\bs r,t) \,=\, \bs \gamma_0(t) \,+\, \bs J \wedge \bs r 
\,-\, \dot \beta \,\bs r ,
\label{eq:gamma}
\end{equation}
where the dot denotes time derivation. Equation (\ref{eq:cond3})
implies that $\partial_t \,(\hbox{\bf curl}\,{\bs \gamma})=\bs 0$ which supplemented
with $\hbox{\bf curl}\,\bs\gamma =\bs J$ [see Eq. (\ref{eq:gamma})], imposes that $\bs J$ be constant
and uniform. It can be shown that $\bs J$ corresponds to the total angular momentum 
of the system, a conserved quantity. In what follows, we will 
put $\bs\gamma_0(t)=\bs 0$ which is always possible up to an innocuous shift of the
velocity origin \cite{comment2}. We focus for simplicity on vanishing angular
momentum solutions, which display already the most interesting properties.
The case $\bs J\neq \bs 0$ is treated in the supplemental material \cite{suppl}.
Combining Eqs.~(\ref{eq:cond3}), (\ref{eq:cond4}) and
(\ref{eq:gamma})%
, we arrive at 
\begin{equation}
\dot\beta \, \left(
2 + \bs r \cdot\! \bs\nabla_{\bs r} 
\right) \,V(\bs r) \,+\, \, 2\beta\, \frac{\partial V}{\partial t}
 \,+\,\dddot \beta \, \frac{r^2}{2} \,=\, 0,
\label{eq:interm}
\end{equation}
up to an irrelevant time dependent function, which can be absorbed into
$V$ without changing the resulting force $\bs F$.
The general solution with $\dot\beta\neq 0$ reads 
\begin{subequations}
\begin{align}
&V(\bs r) = \frac{1}{2} \omega^2(t) r^2 + \frac{b}{r^2},
\label{eq:newsola}\\
 \hbox{with}\quad  & \dddot \beta + 4 \omega^2 \dot \beta \,+\, 
 4 \,\omega \, \dot\omega \,\beta\,=\, 0.
\label{eq:newsolb}
\end{align}
\end{subequations}
Before discussing the possibilities opened by this class 
of solutions, a few words are in order on the static confinement case
($\dot\omega=0$), where it is seen that the breathing mode obeying
$\dddot \beta + 4 \omega^2 \dot \beta =0$ has characteristic frequency 
$2\omega$, twice the trap frequency. Notably, this mode is
unaffected by the non harmonic term in $b/r^2$ (here $b \geq 0$ for
normalizability). While the harmonic solution with $b=0$ 
has been known since the 1870s, the more general form with $b\neq 0$ 
has ben overlooked, and provides a new family of exact solutions to the Boltzmann
equation.

For a general time-dependent confinement ($\dot \omega\neq 0$), Eq.
\eqref{eq:newsolb} gives the evolution of the effective temperature 
$T(t) \propto \beta^{-1}$ \cite{rque50} 
for a given driving of the trap angular frequency, whatever
the collision rate. The evolution of this effective temperature 
is deeply connected to single particle dynamics, which 
allows for the possibility of a parametric resonance \emma{(not shown)}.
In the remainder however, we focus on an inverse perspective: instead
of working out the consequences on dynamical quantities of a given
trap driving $\omega(t)$, we first put forward a desired dynamics for $\beta(t)$,
and find out the required driving in a second step. This strategy is put to work to perform 
on a short time scale the same task as an adiabatic transformation,
which connects two equilibrium states but requires a slow protocol.
Our scheme, which therefore qualifies as a `shortcut to adiabaticity',
can be illustrated on the harmonic case ($b=0$), to 
which we will restrict. It nevertheless also applies for non harmonic
trapping with $b>0$. The idea is to first shape the effective 
temperature to obey a set of boundary conditions, and to design the angular frequency correspondingly. 
In the absence of elastic collisions, an adiabatic change of the strength of confinement obeys the criterion $|\dot{\omega}|\ll \omega^2$, that 
results from the invariance of the one-particle action \cite{LiL92}, and $E/\omega$, where $E$ is the total mechanical 
energy, remains constant. If elastic collisions are at work, the thermodynamical adiabaticity criterion reads 
$|\dot{\omega}|/\omega \ll \tau_{\rm relax}^{-1}$, where $\tau_{\rm relax}$ is the relaxation time needed for the gas to recover equilibrium. 
Under this condition, the population of each single state remains constant as a function of time, and therefore the quantity 
$T/\omega$ remains constant. The relaxation time depends on the relative value between the mean free time $\tau$ and the oscillation period $2\pi/\omega$ \cite{relax,comment4}.

Such an adiabatic evolution can be here easily recovered from Eq.~(\ref{eq:newsolb}) by dropping the $\dddot{\beta}$ term, which yields $\beta(t)\omega(t)=\beta(0)\omega(0)$. 
For this slow evolution, the position-velocity correlation scaling function $\bs\gamma$ vanishes (again, up to a possible rigid rotation).
As demonstrated below, the previously found solutions enable us to generalize the 
concept of shortcut to adiabaticity (STA) for expansions and compressions of a classical gas in 
a potential of the form (\ref{eq:newsola}).
Fast harmonic trap expansions without final excitation were designed 
for single quantum particles using Lewis-Riesenfeldt invariants \cite{Chen1}, 
and for  Bose-Einstein condensates using 
a self-similar ansatz \cite{stabec}.  These expansions have been already successfully implemented with non-interacting thermal atoms
and for Bose-Einstein condensates in the Thomas-Fermi regime \cite{Nice1,Nice2}.

\begin{figure}[htb]
\centerline{\includegraphics[width=7cm]{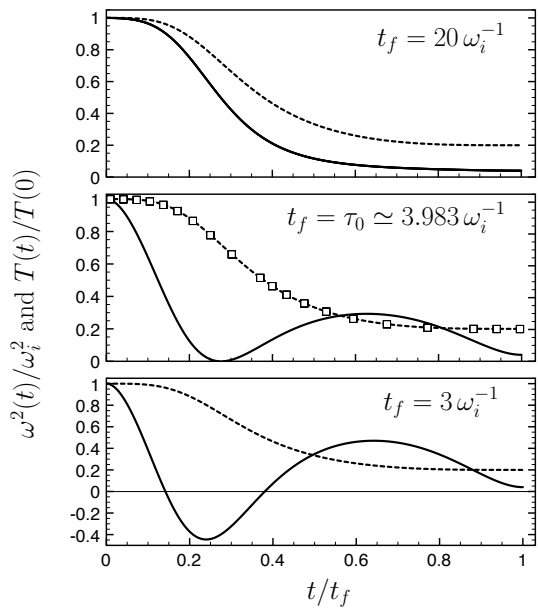}}
\caption{Inverse engineered trap angular frequency as a function of time for a 
decompression $T_f/T_i = 1/5 = \beta_i/\beta_f$ on three different timescales $t_f$ (solid line) 
and the corresponding time evolution of the effective temperature $\beta_i/\beta(s)$ 
(dashed line), where $\beta(s)$ is given given by Eq. \eqref{eq:betades}. 
As explained in the main text, above the critical value 
$t_f>\tau_0\simeq 3.983 \,\omega_i^{-1}$  $\omega^2$ is always positive 
(confining potential) whereas for $t_f<\tau_0$, the trap 
should necessarily be expulsive ($\omega^2<0$) in some intermediate time window.
\emma{For the critical case $t_f=\tau_0$, the square symbols display the temperature
measured in Monte Carlo simulations of a hard disks system, subject to the confining 
time-dependent STA harmonic forcing in $\omega(t)$. They are in excellent agreement with
the desired behavior embodied in Eq. \eqref{eq:betades}, and shown by the dashed line.
A similar agreement with Monte Carlo is found for all transformation times $t_f$.}} 
\label{fig2}
\end{figure}

A remarkable feature of the protocol proposed here for the classical gas is that we can relate two equilibrium states whatever the relaxation time of the system is.
Let us label the initial and final states by $i$ and $f$ respectively: 
$\beta_{i,f}=1/(2k_BT_{i,f})$ and $\bs\gamma_{i,f}=\bs 0$. 
We assume that these states can be related by an adiabatic transformation so that $T_i/\omega_i=T_f/\omega_f$. 
To shape the time dependence of the trap strength and go from one state to the other in an arbitrary time duration $t_f$, we search for a 
polynomial form of $\beta(t)$ that obeys the boundary conditions \cite{selfcons}:
$\beta(0)=\beta_i$, $\dot\beta(0)=0$, $\ddot\beta(0)=0$, $\dddot\beta(0)=0$,  $\beta(t_f)=\beta_f$, $\dot\beta(t_f)=0$, $\ddot\beta(t_f)=0$, and  $\dddot\beta(t_f)=0$. We find
\begin{equation}
\beta(s) = (\beta_f-\beta_i)( - 20s^7+70s^6-84s^5 +35s^4 ) + \beta_i, 
\label{eq:betades}
\end{equation}
with $s=t/t_f$, that varies monotonously from $\beta_i$ to $\beta_f$. Once $\beta(t)$ is known, a first 
order equation on $\omega^2$ (see Eq.~(\ref{eq:newsolb})) remains to be solved with the boundary condition $\omega^2(0)=\omega_i^2$. 
From Eq.~(\ref{eq:newsolb}), we can deduce more on matching conditions: $\dot{\omega}(0)=\dot{\omega}(t_f)=0$. Self-consistency also 
implies that $\omega(t_f)=\omega_f$ \cite{selfcons}. During the evolution, the ratio $T/\omega$ departs from its initial and final values, measuring the deviation from adiabaticity.

As an example, consider a decompression ($\omega_f/\omega_i<1$). On short timescales, a non monotonous variation of $\omega(t)$ is required to fulfill the boundary conditions. 
This occurs with our ansatz for $t_f <  5.904/\omega_i$  when $\omega_f/\omega_i=1/5$. Furthermore, there exists generically a critical time $\tau_0$ for the process duration $t_f$ 
below which $\omega^2(t)$ is negative during some time interval (i.e. the 
potential becomes transiently expulsive, 
\emma{in order to speed up the transformation}). 
For our ansatz and boundary conditions ($\omega_f/\omega_i=1/5$), $\tau_0 \simeq 3.983/\omega_i$. 
Figure \ref{fig2} shows the inverse engineered angular frequency, $\omega(t)$, for the three situations $t_f>\tau_0$, $t_f=\tau_0$ and $t_f<\tau_0$. 
In the case where $T_f>T_i$ (compression), a very similar phenomenology is observed. 
In the slow, adiabatic limit $t_f\to\infty$, we recover an evolution 
with $T \propto \omega$, as it should.

To gain more insight into the transient dynamics, it is instructive
to study the scaling properties of $f(\bs r,\bs v,t)$. It can be noted that
$\beta^{-1/2}$ sets the relevant velocity scale and that conversely 
$\beta^{1/2}$ measures the pertinent length scale. The fact that the product of both
is in $\beta^0$, time independent, can be viewed as a byproduct of angular momentum 
conservation. 
Then, rescaling velocities with respect to the local center-of-mass velocity
[$-\bs \gamma/(2\beta)$, which is position and time dependent], and defining
\begin{equation}
\widetilde{\bs v} = \sqrt{\beta} {\bs v} + \frac{\bs \gamma}{2\sqrt{\beta} },
\quad 
\widetilde{\bs r} = \frac{\bs r}{\sqrt{\beta}},
\end{equation}
the joint distribution of rescaled coordinates 
$\widetilde f(\widetilde{\bs r},\widetilde{\bs v})$ is time
independent \cite{suppl}.
The density of molecules $n(\bs r,t)\equiv \int d\bs v f(\bs r,\bs v,t)$
shares the same feature: when expressed as a function 
of rescaled distance $\widetilde{\bs r}$, it becomes time-independent. 
In the shortcut to adiabaticity protocol, this implies that 
$n(\bs r)$ is of the form
\begin{equation}
n(\bs r,t) \, \propto \, \exp\left( -\frac{\omega_i^2 \beta_i^2}{\beta(t)} \, r^2
\right) ,
\end{equation}
which is exactly the evolution followed under adiabatic transformation. In other words, 
even if transiently expulsive traps are necessary to achieve the transformation
on a time $t_f<\tau_0$, the density remains Gaussian at all times.

As `shortcut to adiabaticity' solutions belong to the kernel of the collision integral, the transformation that 
relates the two thermodynamical equilibrium states can be performed on an arbitrary 
short timescale, irrespective of the collision rate! 
\emma{Yet, the fact that our solutions in the static confinement case do emerge at long
times, might shed doubts on their stability under dynamic and quickly changing
confinement, and thus on the 
existence of the STA route. To address this question, we have implemented Monte Carlo simulations
of a two dimensional hard disks system. They provide the numerical solution to the Boltzmann 
equation (see the Supplemental Material \cite{suppl}). Not only do they back our predictions for static confinements,
but, more importantly, they fully confirm the existence and relevance of the STA route (see the square symbols
in the middle panel of Fig. \ref{fig2}, showing a measured $T(t)$ in remarquable agreement
with the target evolution of Eq. \eqref{eq:betades} \cite{suppl}).
In practice though,} there may be limits to the STA protocol, such as 
$t_f>\tau_0$ if the repeller (expulsive) configuration cannot be implemented. More generally 
short process times imply a growth of the transient energies, both kinetic and potential. The experimental 
constraints on these quantities may impose thus lower limits to $t_f$. For our polynomial ansatz, a harmonic potential, and $\beta_f\gg\beta_i$, 
the transient energies  scale as  $\sim 1/(t_f^{2} \omega_f^{2}\beta_f)$,  or $\sim \hbar N_f/(\omega_f t_f^2)$, $N_f$ being an effective (average) quantum oscillator level number.   
This is the same type of behavior found for single particle expansions \cite{energy} 
and quantifies the third principle, limiting  the speed with which low temperatures may be approached
with the finite energy resources available \cite{energy,constantcomment}.  

The shortcut strategy can be similarly implemented for a harmonically trapped gas in the hydrodynamic 
regime \cite{hydro}. Indeed in this case, the exact scaling solution can be used 
in a similar way as for a Bose-Einstein condensate \cite{stabec}.
Moreover, the previous discussion based on the ansatz (\ref{eq:scaling}) can also be extended to the 
quantum Boltzmann equation with a slightly modified form \cite{suppl},
\begin{equation}
f({\bs r},{\bs v},t) \,=\left(\epsilon+e^{-\alpha({\bs r}, t)-\beta(t)v^2-{\bs \gamma} \cdot {\bs v}+\mu}\right)^{-1},
\end{equation}
where $\epsilon=+1$ for fermions and $\epsilon=-1$ for bosons. One can readily check that this ansatz is in the kernel of the quantum collision 
integral that contains the bosonic amplification or fermionic inhibition factors \cite{baym}, and that the coefficients $\alpha$, $\beta$ and 
$\bs\gamma$ obey the same set of equations as in the classical case, since the ansatz relies on collisional invariants. 

For completeness, we precise that
in the presence of an anisotropic harmonic trap, the breathing mode is coupled to quadrupole modes as experimentally reported in Refs.~\cite{quadupolemonopole0,quadupolemonopole1,quadupolemonopole2} using magnetically trapped samples of cold Bose gases. 
However, in the case of a cylindrically harmonic trap with a large ratio between the transverse and longitudinal angular frequencies ($\omega_\perp \gg \omega_z$), the transverse breathing mode is long-lived since it is only weakly coupled to the longitudinal 
degree 
\cite{monopole2D}.


In conclusion, whereas the derivation of exact solutions to the Boltzmann equation
usually requires some simplifications --a route leading in particular to the so-called 
Maxwell models, or variants thereof \cite{Ernst81}-- we have here explicitly constructed
a family of distribution functions $f({\bs r},{\bs v},t)$ that hold for all intermolecular
(binary) forces. Momentum and energy conservation indeed dictates, through the $H$-theorem,
the form of distributions that nullify the collisional integral $I_{\text{coll}}$ in 
Eq.~(\ref{eq:BE}), and we proceeded by enforcing the consistency of the resulting form 
(\ref{eq:scaling}) with invariance under free streaming. In doing so,
it appears that non-trivial and undamped exact solutions kindred to breathing or expanding behavior
do exist for an external potential of the type (\ref{eq:newsola}). 
We determined from a reverse engineering procedure
what time-dependent harmonic confining frequency was required to achieve a 
fast prescribed time evolution of the system's state.
Ensuing shortcuts to adiabaticity avoid the shortcoming of usual protocols which are performed
slowly to avoid excitations of the final state. This often results in an unacceptably large duration of the experiment,
because of the perturbing effect of noise or the need to repeat the process 
many times, as in atomic clocks.  In addition to the possibility of gas manipulation
on a short time scale, we have emphasized that our protocol applies
to interacting systems, at variance with alternative procedures \cite{deltakick}. 

Interestingly, a breathing mode can be found in larger classes of interacting gases whose collisions cannot be simply described by the $I_{\rm coll}[f,f]$ term. For instance, in two dimensions and for long range interactions of the form $V(\bs r_i, \bs r _j)Ê\propto |\bs r_i- \bs r _j|^{-2}$, an exact scaling solution is found as a result of a hidden symmetry \cite{pitaevskii}. A similar solution holds for 
strongly interacting quantum gases whose collisions are described by the unitary limit in a three-dimensional and isotropic harmonic potential \cite{castin}. As a result, the shortcut to adiabaticity techniques can be also adapted to these interacting systems.

Possible extensions of the present work include the study of non-conservative
force fields, mixture of different molecular species \cite{mixt}
together \emma{with understanding the noteworthy stability of the solutions brought 
to the fore, evidenced by our numerical analysis. 
Another relevant perspective is to take advantage of our dissipationless solutions, confronted to quantum gases experiments, to probe subtle and elusive effects of collisions, such as their coherence that
produces an extra mean-field potential in the Boltzmann description \cite{Snider,david02}
}


We thank E. Torrontegui for useful comments.
We acknowledge financial support from the Agence Nationale pour la Recherche, the 
R\'egion Midi-Pyr\'en\'ees, the university Paul Sabatier (OMASYC project), and the 
NEXT project ENCOQUAM. 
M,J.R.M. acknowledges funding of MCINN, through Project FIS2011-24460 
(partially financed by FEDER funds).
J.G.M. acknowledges funding by projects No. IT472-10, FIS2012-36673-C03-01,
and UFI 11/55.


\begin{thebibliography}{99}
\bibitem{comment3}
The quantity $-\bs\gamma/(2\beta)$ is the local 
mean velocity at 
$\bs r$.
\bibitem{comment}
Adiabaticity is understood here in its `slow enough' quantum mechanical sense,
distinct from the thermodynamic meaning of absence of heat transfer
with the environment.
\bibitem{rque50}
More specifically, we have here $T(t) = [2 k_B \beta(t)]^{-1}$
where $k_B$ is Boltzmann constant.
\bibitem{comment4}
For a harmonically trapped gas 
in equilibrium at a temperature $T_0$, the collisions rate reads $1/\tau=n(0)v_{\rm th}\sigma_0/2$ where $v_{\rm th}=(8 k_B T_0/\pi m)^{1/2}$ 
is the thermal velocity, $n(0)$ the central density and $\sigma_0$ the total elastic cross section.
\bibitem{selfcons}
Equation (\ref{eq:newsolb}) can be recast in ${\rm d}(\omega^2\beta^2)/{\rm d}t=-\beta \dddot{\beta}/2$ which can be integrated directly: 
$(\omega^2\beta^2)(t_f)-(\omega^2\beta^2)(t_i)=[(\beta\ddot{\beta})(t_i)-(\beta\ddot{\beta})(t_f)]/2+[\dot{\beta}^2(t_f)-\dot{\beta}^2(t_i)]/4=0$. 
The last equality is obtained from the boundary conditions and is therefore independent of the specific shape of the 
$\beta(t)$ function that is chosen. We deduce $\omega(t_f)=\omega_i\beta_i/\beta_f=\omega_f$ since we assume that 
the initial and final states can be connected by an adiabatic transformation, and thus obey 
$\omega_i\beta_i=\beta_f\omega_f$.
\bibitem{constantcomment} If the term $b/r^2$ is added to the harmonic potential, 
a $c/(\omega_f t_f^2)$ behavior is still found but $c$ depends on 
$\omega_f \beta_f$ and $b$, and the same conclusion holds.
\end{thebibliography}
\end{document}